\documentclass[aps,twocolumn,showpacs,byrevtex,prl,reprint, nofootinbib]{revtex4-1}
\usepackage{xspace}
\usepackage{epsfig}
\usepackage{dcolumn}
\usepackage{amssymb}   % for math
\usepackage{amsmath}
\usepackage{mathrsfs}
\usepackage{bm}
\usepackage{xcolor}
\usepackage[normalem]{ulem}
\usepackage{graphicx}
\usepackage{subfigure}
\usepackage{siunitx}
\usepackage{pstricks}
\usepackage{multirow}
\usepackage{pst-node}
\usepackage{rotating}
\usepackage{times}
\usepackage{soul}
\usepackage{ulem}
\usepackage[english]{babel}
\addto{\captionsenglish}{%

}
\usepackage[colorlinks,linkcolor=blue, citecolor=blue]{hyperref}
\usepackage{lineno}
\usepackage[T1]{fontenc}
\usepackage{lmodern}
\usepackage{bm}

\newcommand{\EE}{e^+e^-}

\newcommand{\ar}{\rightarrow}
\newcommand{\llb}{\Lambda\bar{\Lambda}}

\newcommand{\bbt}{\bibitem}
\RequirePackage[normalem]{ulem} %DIF PREAMBLE
\RequirePackage{color}\definecolor{RED}{rgb}{1,0,0}\definecolor{BLUE}{rgb}{0,0,1} %DIF PREAMBLE
\DeclareSIUnit{\bmm}{\bm{m}}
\DeclareSIUnit{\clight}{\textnormal{\text{c}}}

\makeatletter
\def\ps@titlepage{%
  \let\@oddhead\@empty
}
\makeatother

\begin{document}
\normalsize
\parskip=5pt plus 1pt minus 1pt
\title{\protect\boldmath Measurement of Born cross section of $e^+e^-\to\Sigma^0\bar{\Sigma}^0$ at $\sqrt{s} = 3.50-4.95$ {\textbf{GeV}}}
%\author{\small\input{authorlist_2024-06-28}}
\author{BESIII Collaboration}
%\date{\today}

%\linenumbers
\begin{abstract}
Using $e^+e^-$ collision data collected with the BESIII detector at the BEPCII collider at 32 center-of-mass energies from 3.50 to 4.95 GeV, corresponding to an integrated luminosity of {25} $\rm{fb^{-1}}$, we measure the Born cross section of the $e^+e^-\to\Sigma^0\bar{\Sigma}^0$ reaction and the effective form factor for the first time.
No significant charmonium(-like) state, i.e., $\psi(3770)$, $\psi(4040)$, $\psi(4160)$, $\psi(4230)$, $\psi(4360)$, $\psi(4415)$, or $\psi(4660)$, decaying into the $\Sigma^0\bar{\Sigma}^0$ final state is observed  by fitting  the $e^+e^- \to \Sigma^0\bar{\Sigma}^0$ dressed cross section. The upper limits for the product of the branching fraction and the electronic partial width at the 90\% confidence level are provided for each assumed charmonium(-like) state for the first time.
In addition, the ratios of the Born cross section and the effective form factor between the $e^+e^-\to\Sigma^0\bar{\Sigma}^0$ and the $e^+e^-\to\Sigma^+\bar{\Sigma}^-$ reactions are provided, which can be used to validate the prediction of the vector meson dominance model.
\end{abstract}
%\pacs{13.20.Gd,13.30.-a, 14.20.Pt}
\maketitle
%\linenumbers
The study of the decays of charmonium(-like) states produced in $e^+e^-$ annihilations into baryon pairs is a key element to test quantum chromodynamics (QCD). The potential model~\cite{Barnes:2005pb} predicts six vector charmonium states in the energy region from 3.7 to 4.7 GeV, identified as the $1D$, $3S$, $2D$, $4S$, $3D$, and $5S$ states~\cite{Farrar}. In the past decades, an abundance of charmonium vector states has been observed at $e^+e^-$ colliders above the open-charm threshold. 
Three conventional charmonium states, i.e., $\psi(4040)$, $\psi(4160)$, and $\psi(4415)$~\cite{BES:2001ckj}, have been observed in open-charm final states; more recently, three nonconventional charmoniumlike states, i.e., $\psi(4230)$, $\psi(4360)$, and $\psi(4660)$, have been observed in hidden-charm final states via initial state radiation (ISR) processes at {\it BABAR} and Belle~\cite{BaBar:2005hhc, Belle:2007dxy, BaBar:2006ait, Belle:2007umv, Belle:2008xmh, BaBar:2012hpr, Belle:2014wyt, BaBar:2012vyb, Belle:2013yex}, or by direct production processes at CLEO~\cite{CLEO} and BESIII~\cite{BESIIIAB, BESIII:2023c.m.v,BESIII:2024ths}.
These states cannot be classified as resonances consisting solely of a $c\bar{c}$ quark pair.
The overpopulation of structures and the discrepancies between potential model predictions and experimental measurements suggest that some of these structures may be candidates for exotic states. To explain their nature~\cite{Chen:2016qju}, many hypotheses including hybrid states~\cite{Briceno}, multiquark states~\cite{Close:2005iz}, and molecular states~\cite{WangQ:2014cvms} have been proposed. However, no definitive conclusion has been drawn. References.~\cite{Wang:2019mhs,Qian:2021neg} suggest that these states are regarded as pure charmonium states if their baryonic decays can be observed.
This situation indicates the imperfect knowledge of the strong interaction, particularly in its nonperturbative aspects. For a better understanding,
more experimental information is desired, such as from the study of charmonium(-like) states decaying into $B\bar{B}$ pairs, where $B$ stands for a baryon, since they provide clear insights into the underlying interaction mechanisms due to their straightforward topology
via three gluons process, one virtual photon process and mixed two gluons and one virtual photon process~\cite{BaldiniFerroli:2019abd}.
However, the studies of the decays of charmonium(-like) states above open-charm threshold into $B\bar B$ pairs are sparse.  Experimental studies have been performed in this energy region by the BESIII experiment~\cite{Ablikim:2013pgf,BESIII:2021ccp, BESIII:2021aer, Ablikim:2019kkp, BESIII:2023rse, BESIII:2017kqg, BESIII:2021cvv, BESIII:2023rse, BESIII:2023rwv,BESIII:2024umc,BESIII:2024ogz, BESIII:2024ues}, and the evidence of only $\psi(3770)\to\llb$ and $\psi(3770)\to\Xi^-\bar\Xi^+$~\cite{BESIII:2021ccp, BESIII:2023rse} processes has been reported, while no significant $B\bar{B}$ decays for other vector charmonium(-like) states have been found. 

Additionally, the measurement of the electromagnetic and effective form factors of baryon resonances is important to explore their internal composition and electric charge distribution. 
According to Ref.~\cite{Dobbs:2014ifa-1}, the branching fractions of charmonium(-like) states decaying into $B\bar B$ are expected to be negligible if the reaction is assumed to be dominated by the nonresonant electromagnetic contribution.
However, BESIII experiment~\cite{BESIII:2021ccp, BESIII:2023rse} reported branching fraction values which are at least 1 order of magnitude larger than this prediction $(\sim10^{-7})$ based on a scaling from the electronic branching fraction values using Eq.~(1) in Ref.~\cite{Dobbs:2014ifa-1}.
The measurement of the cross section of the $e^+e^-\to \Sigma^0\bar{\Sigma}^0$ reaction at center-of-mass energies above the open-charm threshold offers the opportunity to search for the $B\bar B$ charmless decays of the vector charmonium(like) states. Moreover, as proposed by Refs.~\cite{BCS:+0-, Dai:2023vsw}, the measured ratios of the Born cross section and the effective form factor between the $e^+e^- \to \Sigma^0\bar\Sigma^0$ process and its isospin partner  $e^+e^- \to \Sigma^+\bar\Sigma^-$ are important to validate the predictions based on the vector meson dominance model~\cite{vmdmodel,Iachello:1972nu,Iachello:2004aq,Bijker:2004yu, Yang:2019mzq,Li:2021lvs}.

In this paper, we present the measurements of the Born cross section and the effective form factor
for the reaction  $e^+e^- \to \Sigma^0\bar{\Sigma}^0$,
in the range of center-of-mass (c.m.) energy ($\sqrt{s}$) from 3.50 to 4.95 GeV%\cite{BESIII:2015zbz, BESIII:2020eyu},
using $e^+e^-$ collision data corresponding to a total integrated luminosity of 25 $\rm fb^{-1}$ collected with the 
BESIII detector~\cite{besiii} at the BEPCII collider~\cite{BEPCII}. 
The potential resonances are studied by fitting the dressed cross section of the $e^+e^- \to \Sigma^0\bar{\Sigma}^0$ reaction.
The upper limits of products of branching fractions and electronic partial widths at the 90\% CL
for $\psi(3770)$, $\psi(4040)$, $\psi(4160)$, $\psi(4230)$, $\psi(4360)$, $\psi(4415)$, or $\psi(4660)$ decaying into $\Sigma^0\bar{\Sigma}^0$ are provided. In addition, the ratios of  the Born cross section and effective form factor between the $e^+e^-\to\Sigma^+\bar{\Sigma}^-$ and $e^+e^-\to\Sigma^0\bar{\Sigma}^0$ reactions are obtained.

Candidate events of $\EE\to\Sigma^0\bar{\Sigma}^0$ are fully reconstructed, i.e., both the baryon and the antibaryon are reconstructed through the $\Sigma^0 \to \Lambda\gamma$ and $\bar{\Sigma}^0 \to \bar{\Lambda}\gamma$ decays, where  $\Lambda \to p\pi^-$ and $\bar{\Lambda} \to \bar{p}\pi^+$.
The detection efficiency is determined by Monte Carlo (MC) simulations using a sample of 100,000 events for each c.m. energy point, with a uniform phase space (PHSP) model by the {\sc kkmc} generator~\cite{KKMC} including effects of the beam energy spread and ISR corrections.
The $\Sigma^0$ and $\bar{\Sigma}^0$ decay chains are simulated with the PHSP model by the {\sc evtgen} generator~\cite{evtgen2,EVTGEN}. The BESIII geometric description and the detector response are modeled with a {\sc geant4}-based software package~\cite{GEANT4}.

%Tracking and PID
Charged tracks are reconstructed in the multilayer drift chamber (MDC) within the angular region $|\cos\theta| < 0.93$, where $\theta$ is 
the polar angle with respect to the $z$ axis in the laboratory system, which is the MDC symmetry axis.
At least two positive and two negative charged tracks are required to be reconstructed in the MDC.
%PID
Particle identification~(PID) for charged tracks combines measurements of the energy deposited in the MDC~(d$E$/d$x$) and the flight time in the TOF to form likelihoods $\mathcal{L}(h)~(h=p,K,\pi)$ for each hadron $h$ hypothesis.
Tracks are identified as protons when the proton hypothesis has the greatest likelihood ($\mathcal{L}(p)>\mathcal{L}(K)$ and $\mathcal{L}(p)>\mathcal{L}(\pi)$), while charged pions are identified by requiring that $\mathcal{L}(\pi)>\mathcal{L}(p)$ and $\mathcal{L}(\pi)>\mathcal{L}(K)$.
Events with at least one proton, one antiproton, one positive pion, and one negative pion are kept for further analyses.

%gamma selection
Photons are reconstructed from isolated showers in the electromagnetic calorimeter (EMC). 
The energy deposited in the nearby TOF counter is included to improve the reconstruction efficiency and energy resolution. 
The energies of photons are required to be greater than 25 MeV in the EMC barrel region ($|\cos\theta|<0.8$), and greater than 50 MeV in the EMC end cap ($0.86<|\cos\theta|<0.92$).
Furthermore, the difference between the EMC time and the event start time is required to be within $0 < t < 700$ ns, to suppress electronic noise and energy deposits unrelated to the collision events. 
To eliminate showers from charged tracks, the opening angle between the position of each shower in the EMC and any charged track must be greater than 10 degrees. 
Events with at least two photons are kept for further analyses.

%Lambda reconstruction
The $\Lambda (\bar\Lambda)$ candidates are reconstructed via a secondary vertex fit by looping over the $p\pi^- (\bar{p}\pi^+)$ combinations, where for each candidate the corresponding $\chi^2$ value is required to be less than 500.
To suppress background events, the condition $|M_{p\pi^-(\bar{p}\pi^+)} - m_{\Lambda(\bar{\Lambda})}| \leq$ 5 MeV/${c^2}$ is imposed after optimizing the figure-of-merit (FOM = ${\cal{S}}^{\prime}/\sqrt{{\cal{S}}^{\prime} + B}$). Here, $M_{p\pi^-(\bar{p}\pi^+)}$ is the invariant mass of the $p\pi^-(\bar{p}\pi^+)$ combination, $m_{\Lambda(\bar{\Lambda})}$ is the nominal mass of the $\Lambda(\bar{\Lambda})$ baryon from the Particle Data Group (PDG)~\cite{PDG2020}, ${\cal{S}}^{\prime}$ is the number of signal MC events normalized to the real data, and $B$ is the number of the background events taken from the inclusive MC sample of generic $e^+e^-\to$ hadron events.

\begin{figure}[h]
    %\color{red}
    \centering
            \includegraphics[width=0.5\textwidth]{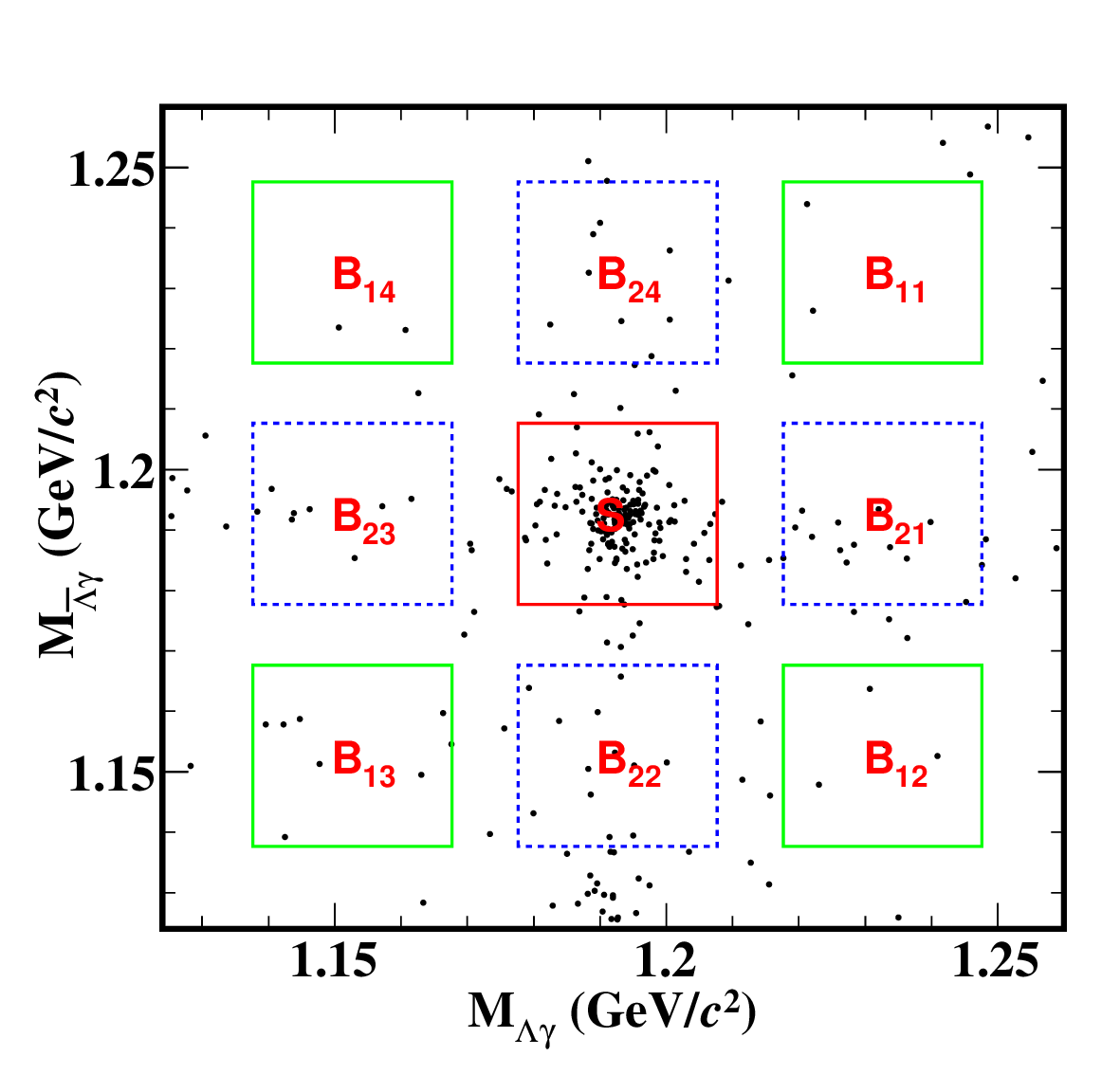}
    \caption{\small {Distribution of  $\rm{M_{\bar{\Lambda}\gamma}}$ versus $\rm{M_{\Lambda\gamma}}$ of accepted candidates in data summed over all energies. The red box represents the signal region and the green boxes and the blue dashed-boxes represent the selected sideband regions.}}
    \label{eachdata}
\end{figure}

%4C
Afterward, a four-constraint (4C)
kinematic fit is applied to all the $\Lambda\bar{\Lambda}\gamma\gamma$ combinations by ensuring the conservation of energy and momentum.
The $\Lambda\bar{\Lambda}\gamma\gamma$ combination with the minimum $\chi^2_{\rm 4C}$ is retained, with an additional requirement of $\chi^2_{\rm 4C} < 100$ to further suppress backgrounds. Among the different combinations of $\Lambda\bar\Lambda\gamma_1\gamma_2$, the one with the minimum value of ${\Delta M=\sqrt{\left(M_{\Lambda\gamma_1}-m_{{\Sigma}^0}\right)^2+\left(M_{\bar{\Lambda}\gamma_2}-m_{\bar{\Sigma}^0}\right)^2}}$ is chosen to identify $\Sigma^{0}\bar{\Sigma}^{0}$ events.
Here, $M_{\Lambda\gamma_1(\bar{\Lambda}\gamma_2)}$ is the invariant mass of the $\Lambda\gamma_1(\bar{\Lambda}\gamma_2)$ combination, and $m_{\Sigma^0(\bar{\Sigma}^0)}$ is the nominal mass of the $\Sigma^0(\bar{\Sigma}^0)$  baryon from the PDG~\cite{PDG2020}.
Figure~\ref{eachdata} shows the distribution of $M_{\bar{\Lambda}\gamma}$ versus $M_{\Lambda\gamma}$.
A clear event accumulation around the $\Sigma^0$ nominal mass can be distinguished. 
The $\Lambda\gamma(\bar{\Lambda}\gamma)$ combination is required to fall within the mass window $|M_{\Lambda\gamma(\bar{\Lambda}\gamma)}-m_{\Sigma^0(\bar{\Sigma}^0)}| \leq$ 15~MeV/${c^2}$, determined by the FOM optimization.

After applying the event selection criteria, the remaining background events mainly come from the process $e^+e^-\to\Lambda\bar\Lambda\to p\bar{p}\pi^+\pi^-$, which has a final state similar to the 
signal one. 
To evaluate the background yield in the signal region, eight sideband regions with the same area $B_{ij}$ (where $i =$ 1, 2 and $j =$ 1, 2, 3, 4) are used
for the $M_{\Lambda\gamma}$ and $M_{\bar{\Lambda}\gamma}$ windows 
as shown in Fig.~\ref{eachdata}. The sideband regions are defined in Table~\ref{Sideband:region}.
%i\textcolor{red}{.e.,
%$B_{11}$: [1.217, 1.247] GeV/$c^2$ and [1.217, 1.247] GeV/$c^2$,
%$B_{12}$: [1.217, 1.247] GeV/$c^2$ and [1.137, 1.167] GeV/$c^2$,
%$B_{13}$: [1.137, 1.167] GeV/$c^2$ and [1.137, 1.167] GeV/$c^2$,
%$B_{14}$: [1.137, 1.167] GeV/$c^2$ and [1.217, 1.247] GeV/$c^2$,
%$B_{21}$: [1.217, 1.247] GeV/$c^2$ and [1.177, 1.207] GeV/$c^2$,
%$B_{22}$: [1.177, 1.207] GeV/$c^2$ and [1.137, 1.167] GeV/$c^2$,
%$B_{23}$: [1.137, 1.167] GeV/$c^2$ and [1.177, 1.207] GeV/$c^2$,
%$B_{24}$: [1.177, 1.207] GeV/$c^2$ and [1.217, 1.247] GeV/$c^2$. }
\begin{table}[!htbp]
    \caption{Definitions of the sideband regions.}
    \centering
    \begin{tabular}{lcc} \\\hline \hline
     &$M_{\gamma\Lambda}$ (GeV/$c^2$) &$M_{\gamma\bar\Lambda}$ (GeV/$c^2$) \\ \hline
$B_{11}$ &[1.217, 1.247] &[1.217, 1.247] \\
$B_{12}$ &[1.217, 1.247] &[1.137, 1.167] \\
$B_{13}$ &[1.137, 1.167] &[1.137, 1.167] \\
$B_{14}$ &[1.137, 1.167] &[1.217, 1.247] \\
$B_{21}$ &[1.217, 1.247] &[1.177, 1.207] \\
$B_{22}$ &[1.177, 1.207] &[1.137, 1.167] \\
$B_{23}$ &[1.137, 1.167] &[1.177, 1.207] \\
$B_{24}$ &[1.177, 1.207] &[1.217, 1.247] \\ \hline \hline
    \end{tabular}
    \label{Sideband:region}
\end{table}
The signal yield $N_{\rm obs}$ for the $\EE\ar\Sigma^0\bar{\Sigma}^0$ process at each c.m. energy point is calculated by subtracting the number of background events from the number of events in the signal region,
$N_{\rm obs} = N_{\rm S} - N_{\rm bkg}$, where $N_{\rm S}$ is the number of events in the signal region, and $N_{\rm bkg}$ is the number of background events scaled by the sideband region, i.e., 
$N_{\rm bkg} = \frac{1}{2}\sum^{4}_{i=1} N_{B_{2i}}-\frac{1}{4}\sum^{4}_{j=1} N_{B_{1j}}$.
By using Rolke's 
method~\cite{Lundberg:2009iu}, the uncertainty of $N_{\rm obs}$ and its upper limit are calculated.
The numerical results are summarized in Supplement Material~\cite{SMABCD}.

\begin{figure}[h]
    \centering
	\includegraphics[width=0.49\textwidth]{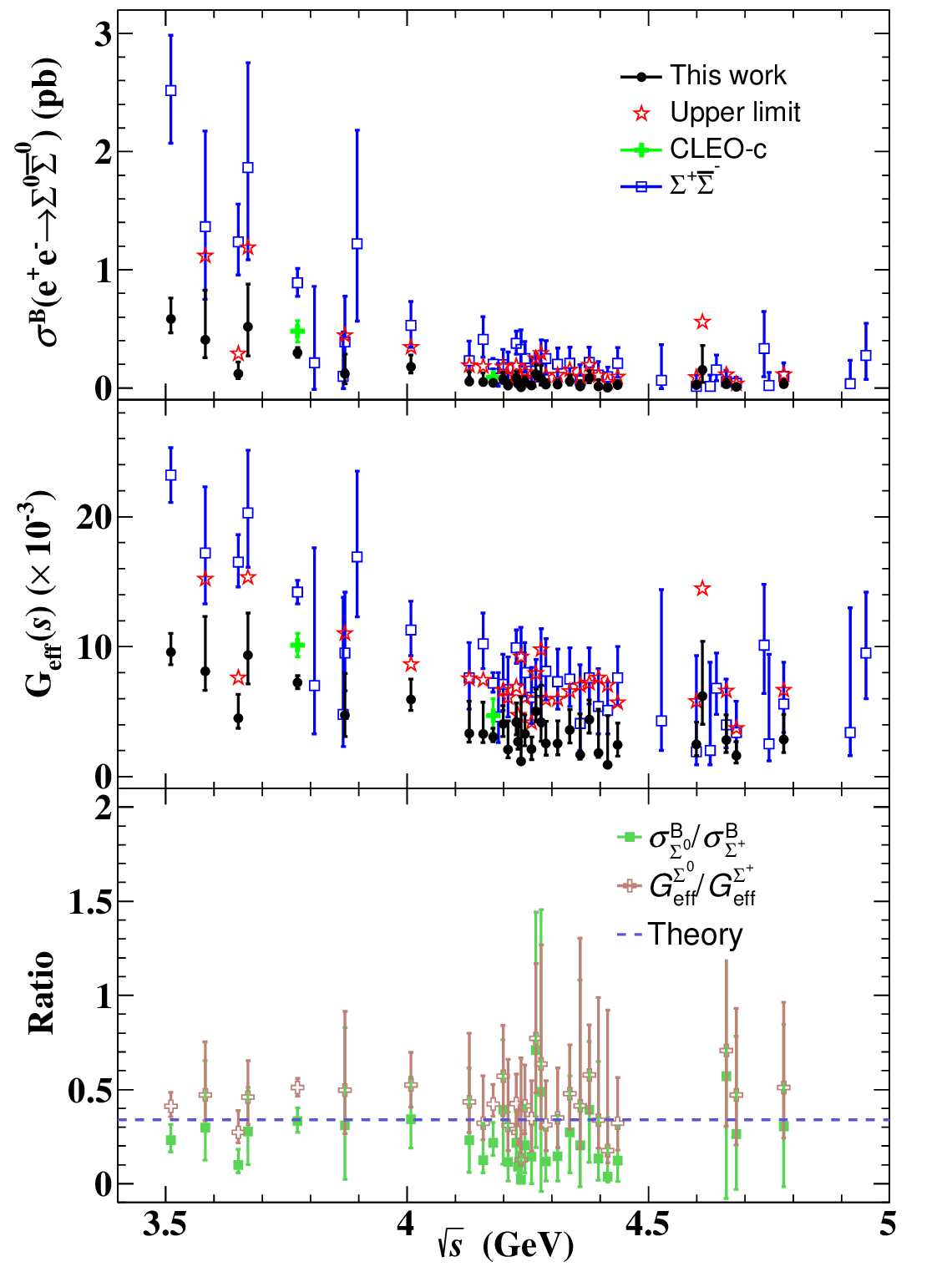}
	\caption{Distributions of the Born cross section (upper plots) and the effective form factor (middle plots) as a function of the c.m. energy for the $e^+e^-\to\Sigma^0\bar{\Sigma}^0$ channel, in comparison to the $e^+e^-\to\Sigma^+\bar{\Sigma}^-$ 
	measurements from BESIII \cite{BESIII:2024umc} and CLEO-c~\cite{Dobbs:2014ifa-1}. The bottom plot shows the ratio of the Born cross section and effective form factor between the results of this work and the BESIII results for $\Sigma^+$ \cite{BESIII:2024umc},  compared to the theoretical predictions~\cite{BCS:+0-}.}
	\label{Fig:ratio_of_sig}
\end{figure}

% \subsection{Determination of Born cross section}
The Born cross section for the $\EE\to\Sigma^0\bar{\Sigma}^0$ process at a given c.m. energy is calculated by
\begin{equation}
\sigma^{B} =\frac{N_{\rm obs}}{{\cal{L}}\cdot(1 + \delta)\cdot\frac{1}{|1 - \Pi|^{2}}\cdot\epsilon\cdot {\cal B}_{\Sigma^0\to \Lambda\gamma}^2 \cdot {\cal B}_{\Lambda\to p\pi^-}^2},
\end{equation}
where ${\cal{L}}$ is the integrated luminosity, $(1 + \delta)$ is the ISR correction factor, $\frac{1}{|1-\Pi|^2}$ is the vacuum polarization (VP) correction factor, $\epsilon$ is the detection efficiency, ${\cal B}_{\Sigma^0\to \Lambda\gamma}$ and ${\cal B}_{\Lambda\to p\pi^-}$ are the corresponding PDG branching fractions~\cite{PDG2020}. The ISR correction factor is obtained using the QED calculation as described in Ref.~\cite{Kuraev:1985hb}. The VP correction factor is calculated according to Ref.~\cite{Jegerlehner:2011ti}. 
The result of the measured Born cross section for each c.m. energy point
is summarized in the Supplement Material~\cite{SMABCD}. 
% are summarized in table~\ref{tab:signal:yields:DD}, and the details are provided in Supplement Material~\cite{SMABCD}. 
Note that 
the efficiencies and ISR correction factors are obtained through an iterative process to accurately measure the Born cross section
as proposed in Ref.~\cite{Sun:2020ehv}.
Figure~\ref{Fig:ratio_of_sig} shows the line shape of
the Born cross section together with the CLEO-c results at $\sqrt{s}$ = 3.770 and 4.160 GeV~\cite{Dobbs:2014ifa-1}, and the BESIII result for $e^+e^-\to\Sigma^+\bar\Sigma^-$~\cite{BESIII:2024umc}.

% \subsection{Determination of effective form factor}
Under the assumption that the dominant process for the reaction of $e^+e^-\to\Sigma^0\bar\Sigma^0$ is the one-photon exchange, the effective form factor $G_{\rm eff}(s)$~\cite{symm:wang} is 
defined as
\begin{equation}
        G_{\rm eff}(s) = \sqrt{\frac{3s\sigma^B}{4\pi\alpha^2C\beta(\frac{2m_{\Sigma^0}^2}{s}+1)}},
\end{equation}
where $\alpha$ is the fine structure constant, $\beta = \sqrt{1-\frac{1}{\tau}}$ is the velocity with $\tau = \frac{s}{4m_{\Sigma^0}^2}$, $m_{\Sigma^0}$ is the mass of $\Sigma^0$ \cite{PDG2020}, and the Coulomb factor $C$~\cite{Baldini, Arbuzov} parametrizes the electromagnetic interaction between the outgoing baryon and antibaryon. For neutral baryons, the Coulomb factor is $C=1$.
Figure~\ref{Fig:ratio_of_sig} shows also a comparison of the effective form factor for the processes $e^+e^-\to\Sigma^0\bar{\Sigma}^0$ and $e^+e^-\to\Sigma^+\bar{\Sigma}^-$ obtained in this work and the previous BESIII result for $\Sigma^+$ \cite{BESIII:2024umc}, together with the CLEO-c result for $\Sigma^0$ \cite{Dobbs:2014ifa-1}.

% \section{Systematic uncertainty}
Systematic uncertainties on the measurement of the Born cross section mainly originate from the integrated luminosity, photon and $\Lambda$($\bar{\Lambda}$) reconstruction, $\Sigma^0$($\bar{\Sigma}^0$) mass window,  background, angular distribution, branching fractions, and input line shape.
%Luminosity
The luminosity at each c.m. energy point is measured using Bhabha events, with the systematic uncertainty of 1.0\% \cite{BESIII:2015qfd} below 4.0~GeV, 0.7\%~\cite{BESIII:2022dxl} from 4.0 to 4.6~GeV, and 0.5\%~\cite{BESIII:2022ulv} above 4.6~GeV.
%Photon reconstruction
The systematic uncertainty due to the photon reconstruction is estimated to be 0.5\% for each photon by analyzing the ISR process $\EE\to\gamma\mu^+\mu^-$; the total systematic uncertainty due to photon reconstruction is considered as 1.0\% summing up linearly the contribution from the two photons.
%The $\Lambda(\bar{\Lambda})$ reconstruction
The uncertainty due to the $\Lambda(\bar{\Lambda})$ reconstruction is obtained including the tracking and PID, the mass window, and the decay length uncertainties. This is studied using a control sample of $J/\psi\to pK^-\bar\Lambda + \rm{c.c.}$ with the same method as used in Ref.~\cite{BESIII:2016ssr, BESIII:2016nix,BESIII:2019dve,BESIII:2022mfx,BESIII:2022lsz,BESIII:2023lkg,BESIII:2023euh,BESIII:lambda_rec_eff} as a result, 
the efficiency difference between data and MC simulation is found to be 1.6\% for the $\Lambda$ reconstruction and 1.3\% for the $\bar\Lambda$ reconstruction. The total uncertainty is 2.9\% by adding them linearly.
%The $\Sigma^0(\bar{\Sigma}^0)$ mass window
The uncertainty due to the $\Sigma^0$ mass window is estimated by varying
the nominal requirements by 5 MeV/$c^2$, corresponding to an uncertainty of 1.0\%.
%Background
The uncertainty associated with the background estimation is assessed by shifting the sideband region outward or inward by 5 MeV/$c^2$, by summing up the samples from all the energy points to reduce the statistical fluctuations. The maximum deviation, 0.2\%, is taken as the systematic uncertainty for background.
%Angular distribution
The uncertainty due to the physical model dependence is estimated to be 2.2\% by comparing the efficiency values obtained with the phase space model and using the real angular distribution, by incorporating the $\Sigma^0$ transverse polarization and the spin correlation
based on the control sample  $\psi(3686)\to\Sigma^0\bar\Sigma^0$.
%Branching fraction     
The uncertainty for the branching fractions of $\Lambda\to p\pi^-$ and $\bar\Lambda\to\bar{p}\pi^+$ is 1.6\%, taken from the PDG~\cite{PDG2020}.
%Input line shape
The uncertainty due to the input line shape of the cross section when determining the ISR correction and the detection efficiency consists of two parts. One part is due to the statistical uncertainty of the input line shape of the cross section, estimated through an alternative input cross section line shape based on a simple power-law (PL) function by varying the central value of the fitted parameters by $\pm1\sigma$.
Another part of uncertainty
for the input line shape arises from the resonance
parameters of the assumed charmonium(-like) states, which are fixed according to the PDG values~\cite{PDG2020} in the fit of the cross section, and evaluated by varying the fixed values of the resonance parameters by 1$\sigma$ uncertainty.
Then, the $(1+\delta)\epsilon$ value for each of c.m. energy point is recalculated and the largest change by considering the contribution from the two parts is taken as the systematic uncertainty. The total systematic uncertainty due to the input line shape is calculated to be 2.0\% by combining the contributions in quadrature.

% \section{Fit to the dressed cross section}
The potential resonances in the cross section for the $\EE\rightarrow\Sigma^0\bar{\Sigma}^0$ reaction are studied by fitting the dressed cross section, $\sigma^{\rm dressed} =\sigma^{B}/|1-\Pi|^2$ (including the VP effect), using the least-squares method, where
$\chi^{2} = \Delta X^{T}V^{-1}\Delta X$.
This is done by considering the covariance matrix $V$ and the vector of residuals $\Delta X$ between the measured and fitted cross section. The covariance matrix incorporates both the correlated and uncorrelated uncertainties across different c.m. energies. The systematic uncertainties from luminosity, photon reconstruction, $\Lambda(\bar{\Lambda})$ reconstruction, $\Sigma^0$ mass window, background, the branching fraction are assumed to be fully correlated among different c.m. energies, while the other systematic uncertainties are assumed to be uncorrelated. 

\begin{figure}[!hbpt]
	\begin{center}
        \includegraphics[width=0.225\textwidth]{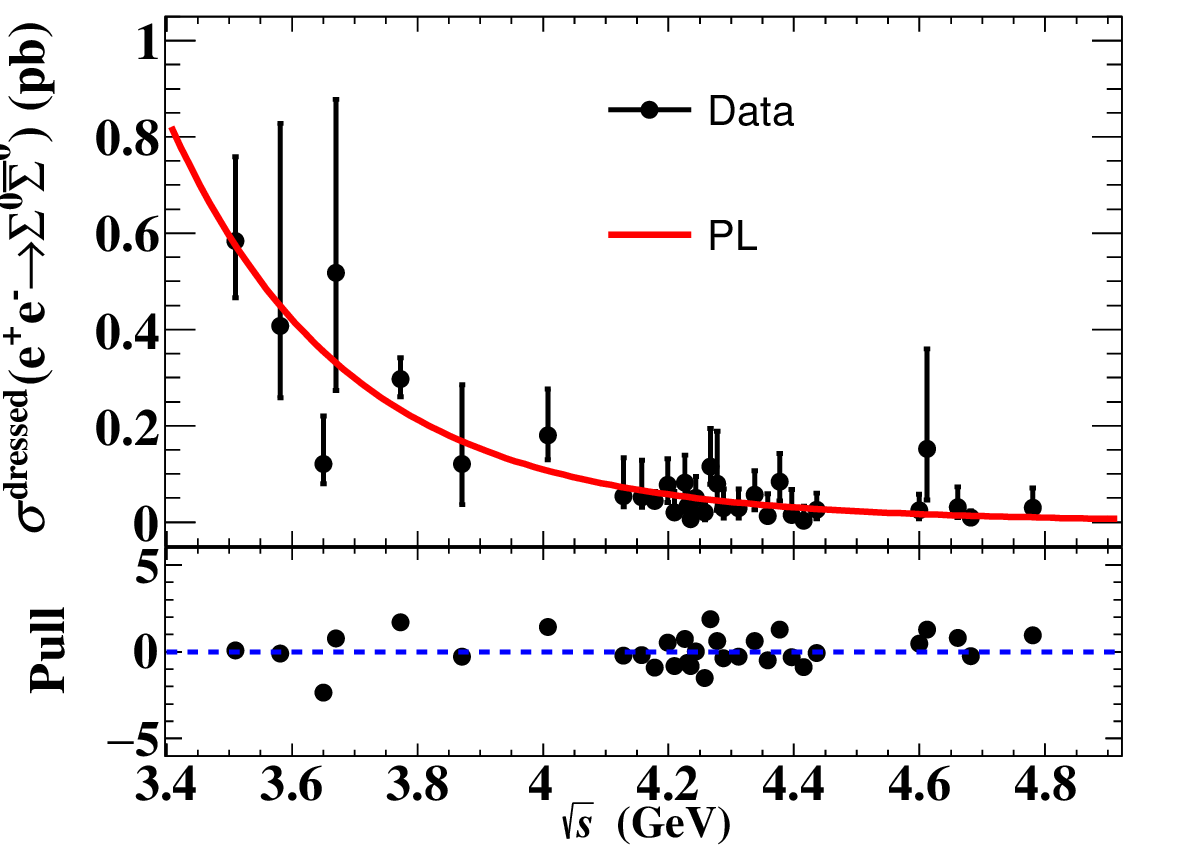}
        \includegraphics[width=0.225\textwidth]{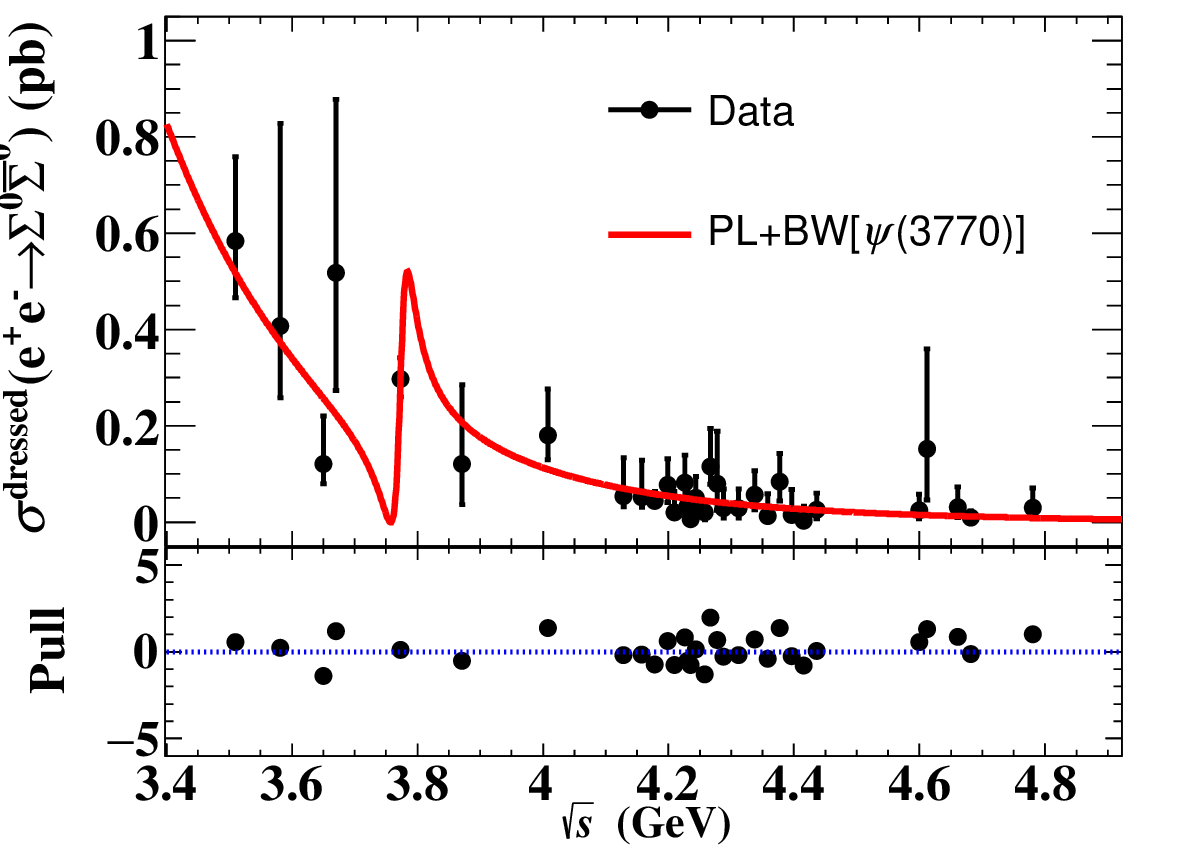}\\
        \includegraphics[width=0.225\textwidth]{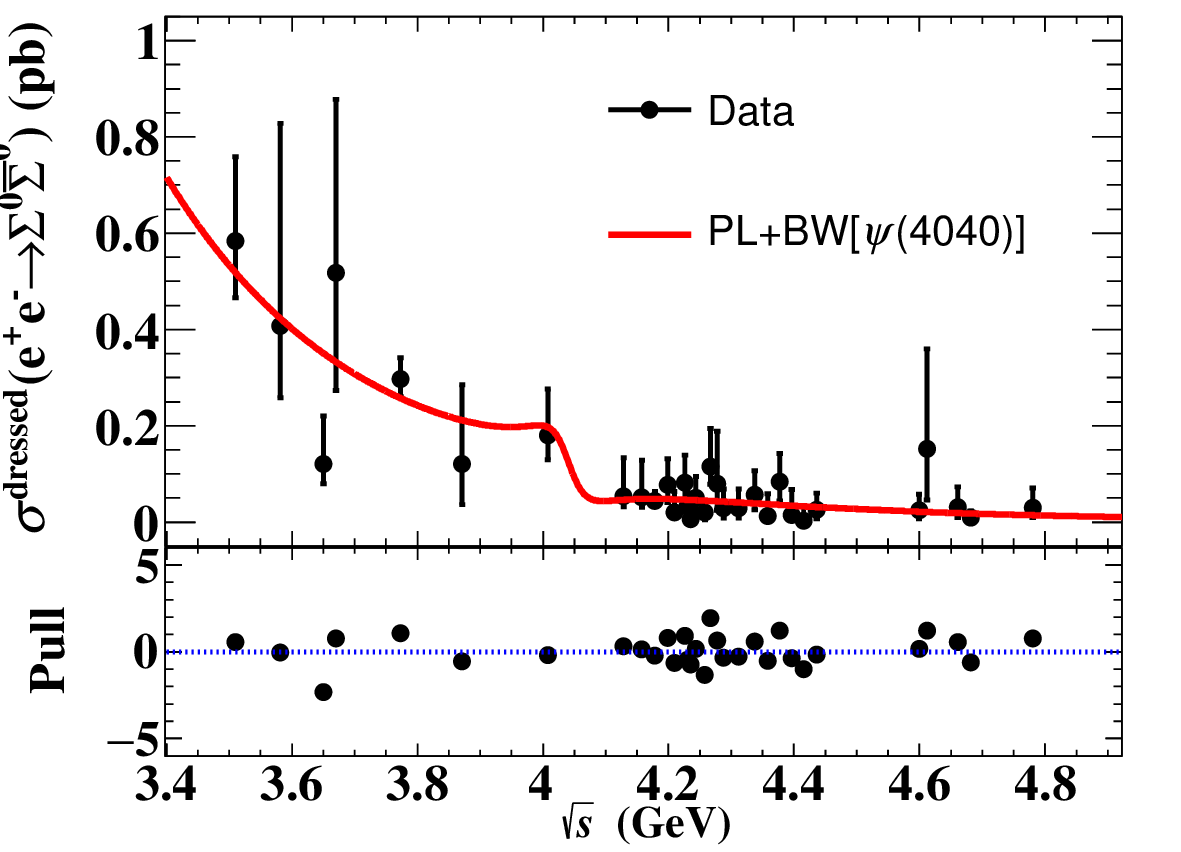}
        \includegraphics[width=0.225\textwidth]{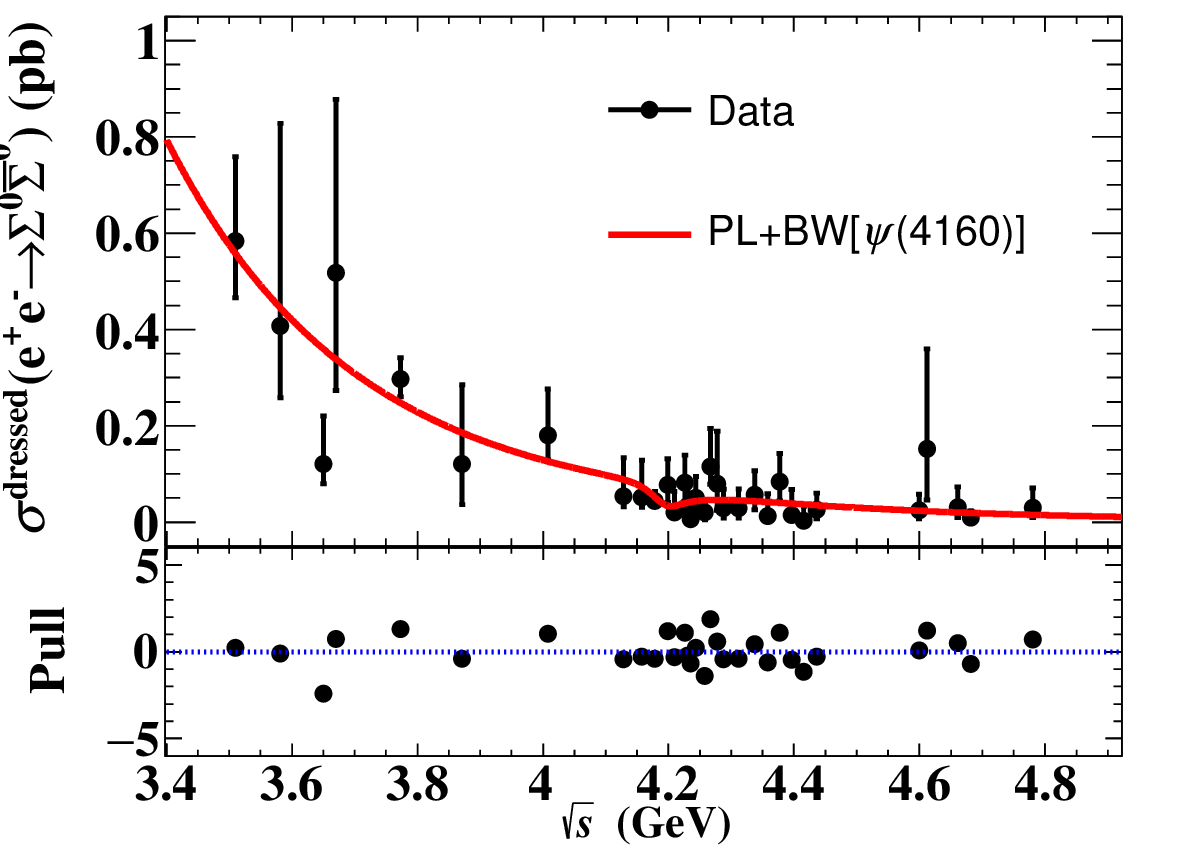}\\
        \includegraphics[width=0.225\textwidth]{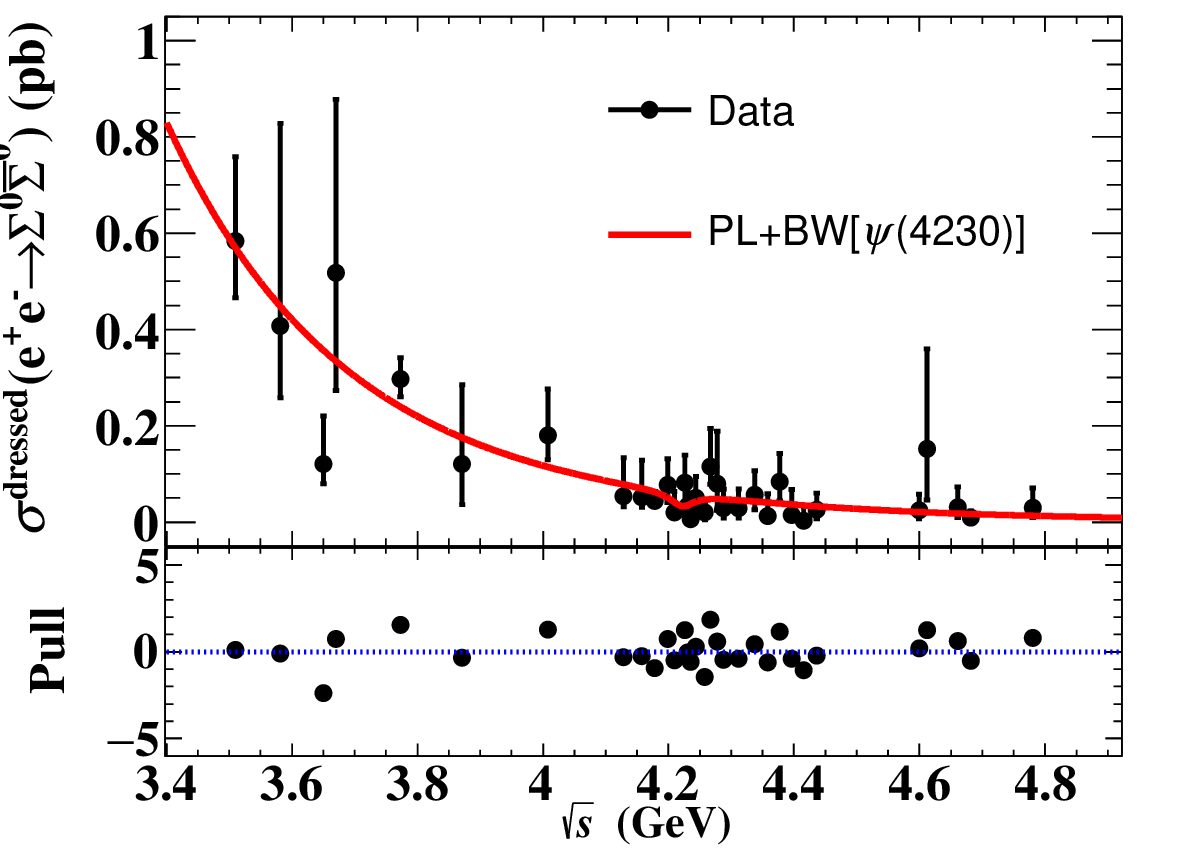}
        \includegraphics[width=0.225\textwidth]{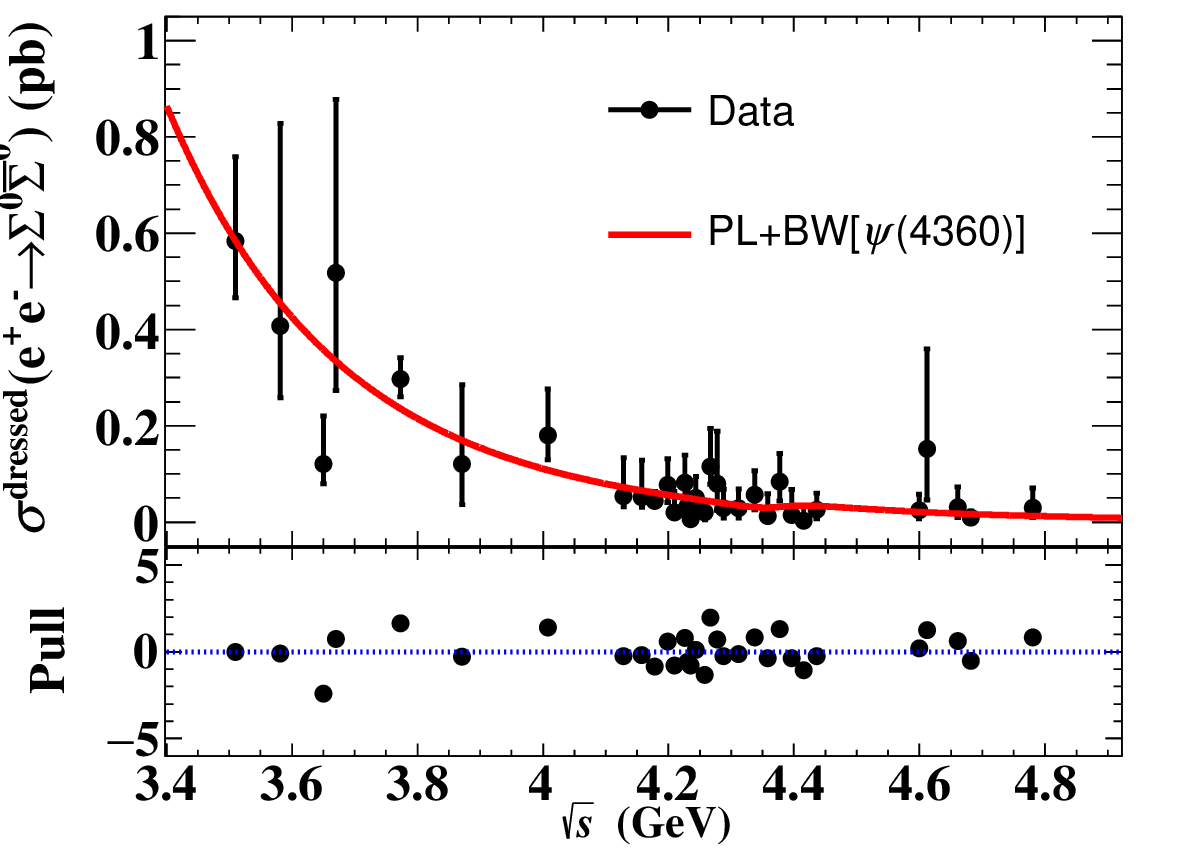}\\
        \includegraphics[width=0.225\textwidth]{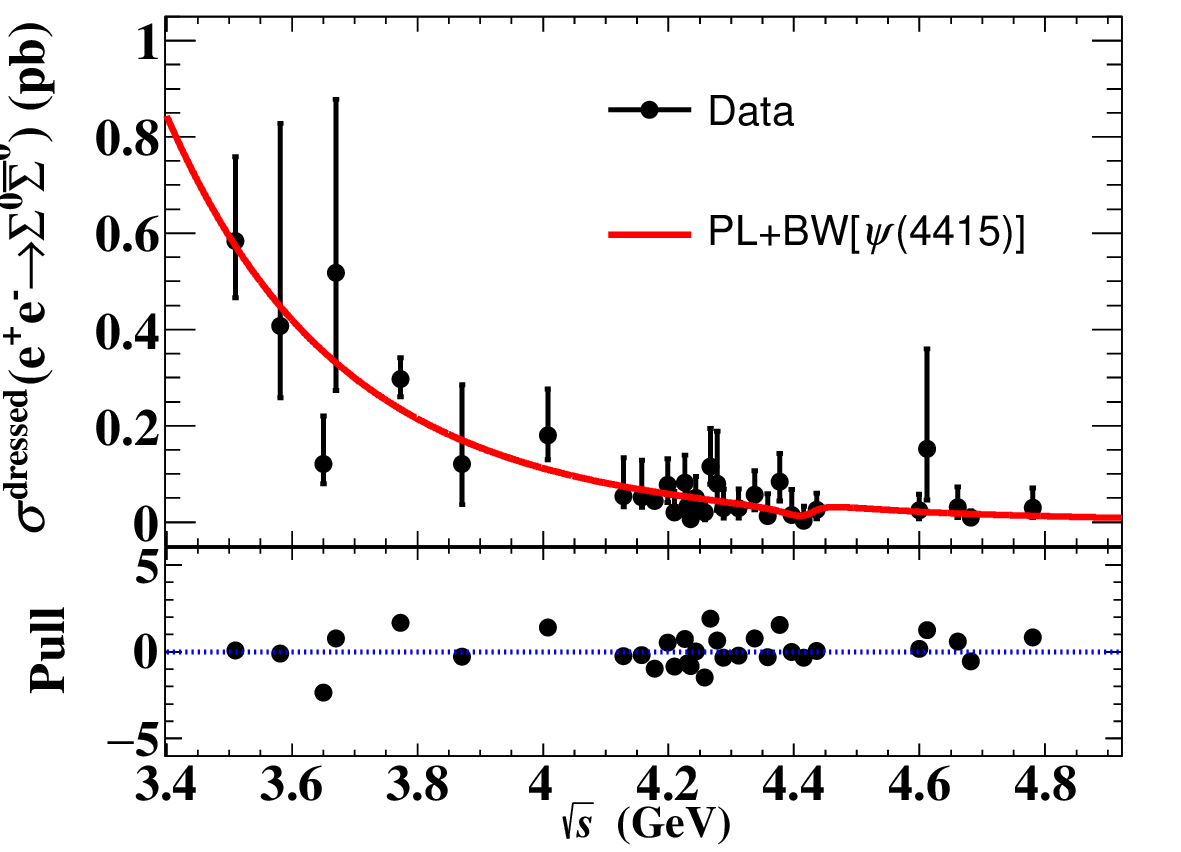}
        \includegraphics[width=0.225\textwidth]{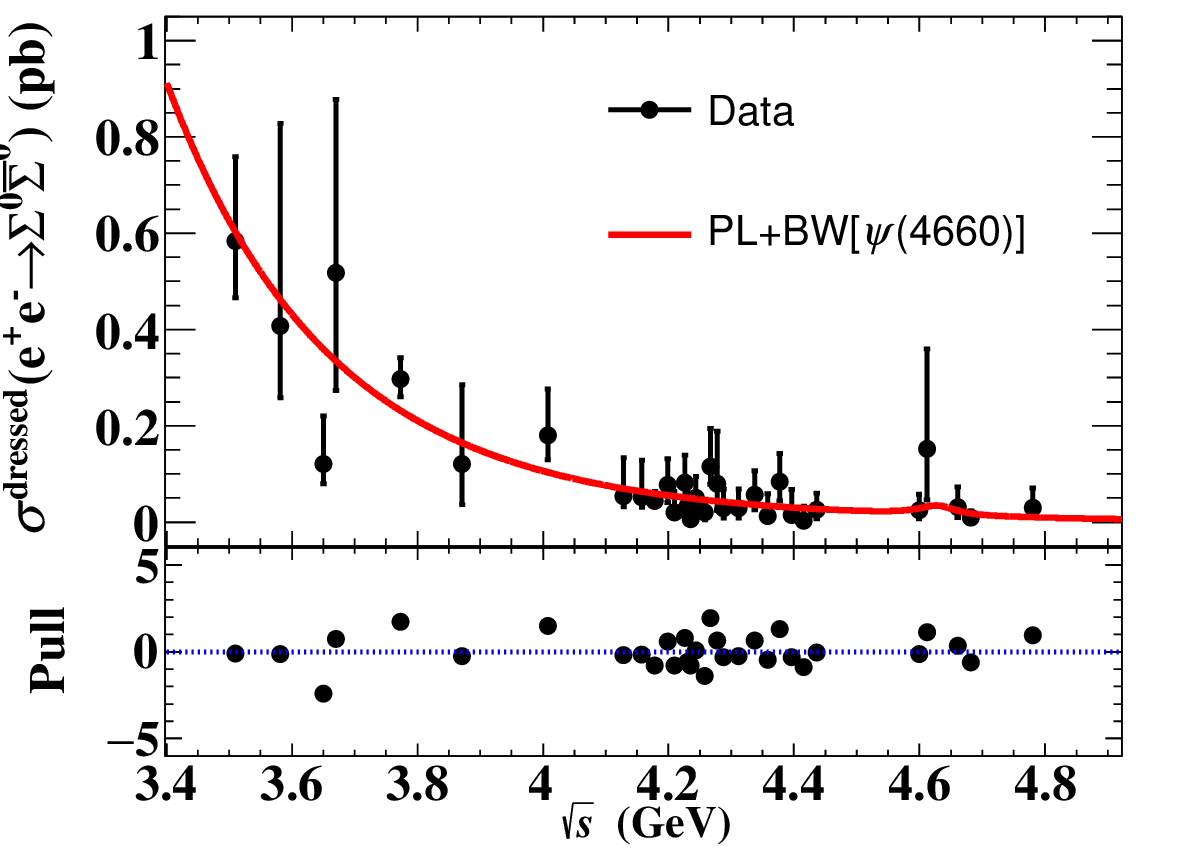}\\	
	\end{center}
	\caption{Fits to the dressed cross section at c.m. energies from 3.510 to \SI{4.951}{GeV} under different assumptions, as indicated in the legend. The dots with error bars represent the dressed cross section and the solid lines indicate the fit results. The error bars consider the statistical and systematic uncertainties summed in quadrature.}
	\label{Fig:XiXi::CS::Line-shape-3773}
\end{figure}

Assuming that the cross section of $\EE\ar\Sigma^0\bar{\Sigma}^0$ includes a resonance [i.e., $\psi(3770)$, $\psi(4040)$, $\psi(4160)$, $\psi(4230)$, $\psi(4360)$, $\psi(4415)$, or $\psi(4660)$] plus a  contribution from the continuum processes, a fit to the dressed cross section is applied using the coherent sum of a power-law function plus a Breit-Wigner (BW) function~\cite{Ablikim:2019kkp}
 \begin{equation}\label{BCS_1}
	\sigma^{\rm dressed}(\sqrt{s})= \left|c_{0}\frac{\sqrt{P(\sqrt{s})}}{\sqrt{s}^{n}} + e^{i\phi}{\rm BW}(\sqrt{s})\sqrt{\frac{P(\sqrt{s})}{P(m)}}\right|^{2}.
 \end{equation}
 Here, $\phi$ is the relative phase between the BW function
  \begin{equation}
{\rm BW}(\sqrt{s}) =\frac{\sqrt{12\pi\Gamma_{ee}{\cal{B}}\Gamma}}{s-m^{2}+im\Gamma}
 \end{equation}
and the PL function, $c_0$ and $n$ are free parameters, $\sqrt{P(\sqrt{s})}$ is the two-body PHSP factor, the mass $m$ and the total width $\Gamma$ are fixed to the PDG values~\cite{PDG2020} of the assumed resonance, and $\Gamma_{ee}{\cal{B}}$ is the product of the electronic partial width and the branching fraction for the resonance decaying into the $\Sigma^0\bar{\Sigma}^0$ final state. 
Figure~\ref{Fig:XiXi::CS::Line-shape-3773} shows the fits to the dressed cross section under different resonance assumptions.
% of $\psi(3770)$, $\psi(4040)$, $\psi(4160)$, $\psi(4230)$, $\psi(4360)$, $\psi(4415)$, and $\psi(4660)$. 
The fit parameters with the assumption of no resonance are $c_0 = 1.8 \pm 0.9$ and $n = 8.5 \pm 0.4$. The resulting parameters under the different resonance assumptions 
%together with their multiple solutions
are summarized in Table~\ref{tab:multisolution}, where the parameters of the PL contributions are omitted for readability. Considering the systematic uncertainties, the significance for each resonance is calculated by comparing the change of $\chi^{2}/n.d.f$ (where $n.d.f.$ is the number of degrees of freedom) with and without the resonance assumption. The dressed cross section is fitted under different assumptions of charmonium(-like) states, i.e., $\psi(3770)$, $\psi(4040)$, $\psi(4160)$, $\psi(4230)$, $\psi(4360)$, $\psi(4415)$, and $\psi(4660)$, one at a time, by using Eq.~(\ref{BCS_1}). No obvious signal is found. The 
$\Gamma_{ee}{\cal{B}}$ and its upper limit including the systematic uncertainty at the 90\% CL, computed using a Bayesian approach \cite{Zhu:2008ca}, are provided. The possible multiple solutions for the resonance parameters in the fit of the dressed cross section are obtained by scanning $\Gamma_{ee}{\cal{B}}$ and $\phi$ in the parameter space, and are provided in the Supplement Material~\cite{SMABCD}.

\begin{table}[h]
    % \color{red}
    % {\color{red}\bf 
    \caption{The fitted resonance parameters for
    $\Gamma_{ee}\mathcal{B}$ $(10^{-3}~\rm{eV})$ and $\phi$ (rad) for two solutions. The fit procedure includes both statistical and systematic uncertainties. Here the $\chi^2/n.d.f$ indicates the fit quality for each assumed resonance. }
    \centering
    \begin{tabular}{l r@{ $\pm$ }l r@{ $\pm$ }l l l}
        \hline
        \hline
        Parameter                                            &\multicolumn{2}{c}{Solution I}         &\multicolumn{2}{c}{Solution II}       &$\chi^2/n.d.f$
        \\
        \hline
        $\phi_{\psi(3770)}$                                    &$-0.8$  &$0.1$                         &$-0.7$   &$0.1$                       &\multirow{2}{*}{$22/(32-4)$}\\$\Gamma_{ee}\mathcal{B}_{\psi(3770)}$            &$17.2$  &$1.4~(\textless 29.2)$      &$14.3$   &$1.1~$                                   \\

        $\phi_{\psi(4040)}$                                    &$-1.9$  &$0.1$                         &$-3.0$   &$0.6$                       &\multirow{2}{*}{$23/(32-4)$}\\$\Gamma_{ee}\mathcal{B}_{\psi(4040)} $            &$37.7$  &$13.9~(\textless 66.1)$      &$3.6$   &$2.3~$                                   \\

        $\phi_{\psi(4160)}$                                    &$-1.7$  &$0.1$                         &$-2.1$   &$0.3$                       &\multirow{2}{*}{$25/(32-4)$}\\$\Gamma_{ee}\mathcal{B}_{\psi(4160)}$            &$17.7$  &$2.4~(\textless 22.7)$      &$0.7$   &$0.7~$                                   \\

        $\phi_{\psi(4230)}$                                    &$-1.6$  &$0.1$                         &$-1.6$   &$0.5$                       &\multirow{2}{*}{$26/(32-4)$}\\$\Gamma_{ee}\mathcal{B}_{\psi(4230)} $            &$10.7$  &$1.8~(\textless 14.7)$      &$0.2$   &$0.3~$                                   \\

        $\phi_{\psi(4360)}$                                    &$-1.6$  &$1.8$                         &$-0.5$   &$1.1$                       &\multirow{2}{*}{$27/(32-4)$}\\$\Gamma_{ee}\mathcal{B}_{\psi(4360)}$            &$20.9$  &$11.5~(\textless 29.3)$      &$0.2$   &$0.5~$                                   \\

        $\phi_{\psi(4415)}$                                    &$-1.5$  &$0.2$                         &$-0.9$   &$0.5$                       &\multirow{2}{*}{$27/(32-4)$}\\$\Gamma_{ee}\mathcal{B}_{\psi(4415)}$            &$7.6$  &$2.7~(\textless 14.0)$      &$0.5$   &$1.0~$                                   \\

        $\phi_{\psi(4660)}$                                    &$-1.7$  &$0.3$                         &$1.7$   &$1.1$                       &\multirow{2}{*}{$27/(32-4)$}\\$\Gamma_{ee}\mathcal{B}_{\psi(4660)}$            &$9.9$  &$3.5~(\textless 17.4)$      &$0.5$   &$1.0~$                                   \\
        \hline
        \hline
    \end{tabular}
    \label{tab:multisolution}
\end{table}
%maybe too long?
% \begin{figure}[!htbp]
% 	\begin{center}
% 	\includegraphics[width=0.225\textwidth]{figure/2dsolscan/shape_likelyhood_3770.root.eps}
% 	\put(-45, 55){\bf (a)}
%     \includegraphics[width=0.225\textwidth]{figure/2dsolscan/shape_likelyhood_4040.root.eps}
%      \put(-45, 55){\bf (b)}\\
%     \includegraphics[width=0.225\textwidth]{figure/2dsolscan/shape_likelyhood_4160.root.eps}
%      \put(-45, 55){\bf (c)}
%     \includegraphics[width=0.225\textwidth]{figure/2dsolscan/shape_likelyhood_4230.root.eps}
%     \put(-45, 55){\bf (d)}\\
%   \includegraphics[width=0.225\textwidth]{figure/2dsolscan/shape_likelyhood_4360.root.eps}
%       \put(-45, 55){\bf (e)}
%     \includegraphics[width=0.225\textwidth]{figure/2dsolscan/shape_likelyhood_4415.root.eps}
%       \put(-45, 55){\bf (f)}\\
%     \includegraphics[width=0.225\textwidth]{figure/2dsolscan/shape_likelyhood_4660.root.eps}
%     \put(-45, 55){\bf (g)}
% 	\end{center}
% 	\caption{The contour of $\Gamma_{ee}\mathcal{B}$ and $\phi$ on the distribution of $\chi^2$ values for (a) $\psi(3770)$, (b) $\psi(4040)$, (c) $\psi(4160)$, (d) $Y(4230)$, (e) $Y(4360)$, (f) $\psi(4415)$, or (g) $Y(4660)$ decaying into the $\Sigma^0\bar{\Sigma}^0$ final states. The red points represent the center values of nominal values from the best fit.
% 	}
% 	\label{Fig:scan}
% \end {figure}

% \section{Summary}
In summary, using $\EE$ collision data collected by the BESIII detector at the BEPCII collider corresponding to an integrated luminosity of  \SI{25}{fb^{-1}}, the Born cross section and the effective form factor for the $e^+e^- \to \Sigma^0\bar{\Sigma}^0$ reaction are measured at 32 c.m. energies ranging from 3.50 to 4.95 GeV for the first time.
Our work provides not only better precision than the CLEO-c measurement at $\sqrt{s}$ = 3.770 and 4.160 GeV~\cite{Dobbs:2014ifa-1}, but also offers measurements at many points over a wide energy range. 
A fit to the dressed cross section of the $\EE\ar\Sigma^0\bar{\Sigma}^0$ reaction is performed under the assumption of a charmonium(-like) resonance [i.e., $\psi(3770)$, $\psi(4040)$, $\psi(4160)$, $\psi(4230)$, $\psi(4360)$, $\psi(4415)$, or $\psi(4660)$] plus a continuum contribution. 
No significant signal is found. 
The products of branching fractions and electronic partial widths for these charmonium(-like) states decaying into the $\Sigma^0\bar{\Sigma}^0$ final state, as well as their upper limits at the 90\% CL, are provided for the first time.
In addition, 
the ratios of Born cross section and effective form factor 
of the $e^+e^-\to\Sigma^0\bar{\Sigma}^0$ reaction over the charged $e^+e^-\to\Sigma^+\bar{\Sigma}^-$ reaction are also provided, and the obtained results agree with the naive 
extrapolation of the theoretical prediction~\cite{BCS:+0-} to higher energies, and thereby can be used to validate the vector meson dominance model for these processes~\cite{Iachello:1972nu,Iachello:2004aq,Bijker:2004yu, Yang:2019mzq,Li:2021lvs}.
These results are important to study the vector charmonium(-like) states decaying into $B\bar{B}$ final states, and further investigate the nature of charmonium(-like) state production above the open-charm threshold. 

\textbf{$Acknowledgement$}\textemdash
 The BESIII Collaboration thanks the staff of BEPCII and the IHEP computing center for their strong support. This work is supported in part by National Natural Science Foundation of China (NSFC) under Contract No. 
No. 12075107, No. 12247101, No. 11635010, No. 11735014, No. 11935015, No. 11935016, No. 11935018, No. 12025502, No. 12035009, No. 12035013, No. 12061131003, No. 12192260, No. 12192261, No. 12192262, No. 12192263, No. 12192264, No. 12192265, No. 12221005, No. 12225509, No. 12235017, No. 12361141819; 
National Key R\&D Program of China under Contract No. 2023YFA1606000, No. 2020YFA0406300, No. 2020YFA0406400;
the Fundamental Research Funds for the Central Universities under Contract No. lzujbky-2024-jdzx06;
the Natural Science Foundation of Gansu Province under Contract No. 22JR5RA389; 
the ‘111 Center’ under Contract No. B20063; the Chinese Academy of Sciences (CAS) Large-Scale Scientific Facility Program; the CAS Center for Excellence in Particle Physics (CCEPP); Joint Large-Scale Scientific Facility Funds of the NSFC and CAS under Contract No. U1832207; 100 Talents Program of CAS; The Institute of Nuclear and Particle Physics (INPAC) and Shanghai Key Laboratory for Particle Physics and Cosmology; German Research Foundation DFG under Contract Nos. FOR5327, GRK 2149; Istituto Nazionale di Fisica Nucleare, Italy; Knut and Alice Wallenberg Foundation under Contract Nos. 2021.0174, 2021.0299; Ministry of Development of Turkey under Contract No. DPT2006K-120470; National Research Foundation of Korea under Contract No. NRF-2022R1A2C1092335; National Science and Technology fund of Mongolia; National Science Research and Innovation Fund (NSRF) via the Program Management Unit for Human Resources \& Institutional Development, Research and Innovation of Thailand under Contract Nos. B16F640076, B50G670107; Polish National Science Centre under Contract No. 2019/35/O/ST2/02907; Swedish Research Council under Contract No. 2019.04595; The Swedish Foundation for International Cooperation in Research and Higher Education under Contract No. CH2018-7756; U. S. Department of Energy under Contract No. DE-FG02-05ER41374.

\begin{widetext}
\begin{center}
\small
M.~Ablikim$^{1}$, M.~N.~Achasov$^{4,c}$, P.~Adlarson$^{76}$, O.~Afedulidis$^{3}$, X.~C.~Ai$^{81}$, R.~Aliberti$^{35}$, A.~Amoroso$^{75A,75C}$, Y.~Bai$^{57}$, O.~Bakina$^{36}$, I.~Balossino$^{29A}$
% ...
\\
\vspace{0.2cm}
(BESIII Collaboration)\\
\vspace{0.2cm} {\it
$^{1}$ Institute of High Energy Physics, Beijing 100049, People's Republic of China\\
% ...
$^{3}$ Bochum  Ruhr-University, D-44780 Bochum, Germany\\
$^{4}$ Budker Institute of Nuclear Physics SB RAS (BINP), Novosibirsk 630090, Russia\\
% ...
$^{29}$ INFN Sezione di Ferrara, (A)INFN Sezione di Ferrara, I-44122, Ferrara, Italy; (B)University of Ferrara,  I-44122, Ferrara, Italy\\
% ...
$^{35}$ Johannes Gutenberg University of Mainz, Johann-Joachim-Becher-Weg 45, D-55099 Mainz, Germany\\
$^{36}$ Joint Institute for Nuclear Research, 141980 Dubna, Moscow region, Russia\\
% ...
$^{57}$ Southeast University, Nanjing 211100, People's Republic of China\\
% ...
$^{75}$ University of Turin and INFN, (A)University of Turin, I-10125, Turin, Italy; (B)University of Eastern Piedmont, I-15121, Alessandria, Italy; (C)INFN, I-10125, Turin, Italy\\
$^{76}$ Uppsala University, Box 516, SE-75120 Uppsala, Sweden\\
% ...
$^{81}$ Zhengzhou University, Zhengzhou 450001, People's Republic of China\\

\vspace{0.2cm}
% ...
$^{c}$ Also at the Novosibirsk State University, Novosibirsk, 630090, Russia\\
% ...
}
\end{center}
\end{widetext}


\begin{thebibliography}{99}

\bibitem{Barnes:2005pb}
T.~Barnes, S.~Godfrey and E.~S.~Swanson,
\textcolor{blue}{\href{https://doi.org/10.1103/PhysRevD.72.054026} {\text{Phys. Rev. D} \textbf{72}, 054026 (2005).}}

\bbt{Farrar} 
N.~Brambilla \textit{et al.}
\textcolor{blue}{\href{https://doi.org/10.1140/epjc/s10052-010-1534-9} {\text{Eur. Phys. J. C} \textbf{71}, 1534 (2011).}} 


\bibitem{BES:2001ckj}
J. Z. Bai \textit{et al.} (BES Collaboration),
\textcolor{blue}{\href{https://doi.org/10.1103/PhysRevLett.88.101802} {\text{Phys. Rev. Lett.} \textbf{88}, 101802 (2002).}}

\bibitem{BaBar:2005hhc}
B.~Aubert {\it et al.} (\textit{BABAR} Collaboration),
\textcolor{blue}{\href{https://doi.org/10.1103/PhysRevLett.95.142001} {\text{Phys. Rev. Lett.} \textbf{95}, 142001 (2005).}}

\bibitem{BaBar:2006ait}
B.~Aubert {\it et al.} (\textit{BABAR} Collaboration),
\textcolor{blue}{\href{https://doi.org/10.1103/PhysRevLett.98.212001} {\text{Phys. Rev. Lett.} \textbf{98}, 212001 (2007).}} 

\bibitem{Belle:2007umv}
G.~Pakhlova {\it et al.} (Belle Collaboration),
\textcolor{blue}{\href{https://doi.org/10.1103/PhysRevLett.99.142002} {\text{Phys. Rev. Lett.} \textbf{99}, 142002 (2007).}} 

\bibitem{Belle:2007dxy}
G.~Pakhlova {\it et al.} (Belle Collaboration),
\textcolor{blue}{\href{https://doi.org/10.1103/PhysRevLett.99.182004} {\text{Phys. Rev. Lett.} \textbf{99}, 182004 (2007).}} 

\bibitem{Belle:2008xmh}
G.~Pakhlova {\it et al.} (Belle Collaboration),
\textcolor{blue}{\href{https://doi.org/10.1103/PhysRevLett.101.172001} {\text{Phys. Rev. Lett.} \textbf{101}, 172001 (2008).}} 

\bibitem{BaBar:2012vyb}
B.~Aubert {\it et al.} (\textit{BABAR} Collaboration),
\textcolor{blue}{\href{https://doi.org/10.1103/PhysRevD.86.051102} {\text{Phys.Rev.D} \textbf{86}, 051102 (2012).}} 

\bibitem{Belle:2013yex}
G.~Pakhlova {\it et al.} (Belle Collaboration),
\textcolor{blue}{\href{https://doi.org/10.1103/PhysRevLett.111.019901} {\text{Phys. Rev. Lett.} \textbf{111}, 019901 (2013).}} 

\bibitem{BaBar:2012hpr}
B.~Aubert {\it et al.} (\textit{BABAR} Collaboration),
\textcolor{blue}{\href{https://doi.org/10.1103/PhysRevD.89.111103} {\text{Phys. Rev. D} \textbf{89}, 111103 (2014).}} 

\bibitem{Belle:2014wyt}
G.~Pakhlova {\it et al.} (Belle Collaboration),
\textcolor{blue}{\href{https://doi.org/10.1103/PhysRevD.91.112007} {\text{Phys. Rev. D} \textbf{91}, 112007 (2015).}} 

\bbt{CLEO}
T.~E.~Coan {\it et al.} (CLEO Collaboration),
\textcolor{blue}{\href{https://doi.org/10.1103/PhysRevLett.96.162003} {\text{Phys. Rev. Lett.} \textbf{96}, 162003 (2006).}} 

\bbt{BESIIIAB}
M.~Ablikim \textit{et al.} (BESIII Collaboration),
\textcolor{blue}{\href{https://doi.org/10.1103/PhysRevLett.114.092003} {\text{Phys. Rev. Lett.} \textbf{114}, 092003 (2015).}} 

\bibitem{BESIII:2023c.m.v}
M.~Ablikim \textit{et al.} (BESIII Collaboration),
\textcolor{blue}{\href{https://doi.org/10.1103/PhysRevLett.130.121901} {\text{Phys. Rev. Lett.} \textbf{130}, 121901 (2023).}} 

\bibitem{BESIII:2024ths}
M.~Ablikim \textit{et al.} (BESIII Collaboration),
%``Precise Measurement of Born Cross Sections for e+e-\textrightarrow{}DD\textasciimacron{} at s=3.80-4.95\,\,GeV,''
\textcolor{blue}{\href{https://journals.aps.org/prl/abstract/10.1103/PhysRevLett.133.081901}
{\text{Phys. Rev. Lett.} \textbf{133}, 081901 (2024).}} 


\bibitem{Chen:2016qju}
H.~X.~Chen, W.~Chen, X.~Liu and S.~L.~Zhu,
\textcolor{blue}{\href{https://doi.org/10.1016/j.physrep.2016.05.004} {\text{Phys. Rept.} \textbf{639}, 1-121 (2016).}}


\bbt{Briceno}
R.~A.~Briceno \text{et al.,}
\textcolor{blue}{\href{https://doi.org/10.1088/1674-1137/40/4/042001} {\text{Chin. Phys. C} \textbf{40}, 042001 (2016).}} 

\bibitem{Close:2005iz}
F.~E.~Close and P.~R.~Page,
\textcolor{blue}{\href{https://doi.org/10.1016/j.physletb.2005.09.016} {\text{Phys. Lett. B} \textbf{628}, 215-222 (2005).}} 


\bibitem{WangQ:2014cvms}
M.~Cleven, Q.~Wang, F.~K.~Guo, C.~Hanhart, U.~G.~Meißner and Q.~Zhao,
\textcolor{blue}{\href{https://doi.org/10.1103/PhysRevD.90.074039} {\text{Phys. Rev. D} \textbf{90}, 074039 (2014).}} 


\bibitem{Wang:2019mhs}
J.~Z.~Wang, D.~Y.~Chen, X.~Liu and T.~Matsuki,
\textcolor{blue}{\href{https://doi.org/10.1103/PhysRevD.99.114003}
{\text{Phys. Rev. D} \textbf{99}, 114003 (2019).}}

\bibitem{Qian:2021neg} 
R.~Q.~Qian, Q.~Huang and X.~Liu, 
\textcolor{blue}{\href{https://doi.org/10.1016/j.physletb.2022.137292} {\text{Phys. Lett. B} \textbf{833}, 137292 (2022).}} 


\bibitem{BaldiniFerroli:2019abd}
R.~Baldini Ferroli, A.~Mangoni, S.~Pacetti and K.~Zhu,
\textcolor{blue}{\href{https://www.sciencedirect.com/science/article/pii/S0370269319307634?via\%3Dihub}
{\text{Phys. Lett. B} \textbf{799}, 135041 (2019).}}



\bibitem{Ablikim:2013pgf} 
M.~Ablikim \textit{et al.} (BESIII Collaboration),
\textcolor{blue}{\href{https://doi.org/10.1103/PhysRevD.87.112011} {\text{Phys. Rev. D} \textbf{87}, 112011 (2013).}} 

\bibitem{BESIII:2017kqg}
M.~Ablikim \textit{et al.} (BESIII Collaboration),
\textcolor{blue}{\href{https://doi.org/10.1103/PhysRevLett.120.132001} {\text{Phys. Rev. Lett.} \textbf{120}, 132001 (2018).}} 

\bibitem{Ablikim:2019kkp} 
M.~Ablikim \textit{et al.} (BESIII Collaboration),
\textcolor{blue}{\href{https://doi.org/10.1103/PhysRevLett.124.032002} {\text{Phys. Rev. Lett.} \textbf{124}, 032002 (2020).}} 

\bibitem{BESIII:2021ccp} 
M.~Ablikim \textit{et al.} (BESIII Collaboration),
\textcolor{blue}{\href{https://doi.org/10.1103/PhysRevD.104.L091104} {\text{Phys. Rev. D} \textbf{104}, L091104 (2021).}} 


\bibitem{BESIII:2021aer}
M.~Ablikim \textit{et al.} (BESIII Collaboration),
\textcolor{blue}{\href{https://www.sciencedirect.com/science/article/pii/S0370269321004974?via\%3Dihub}{\text{Phys. Lett. B} \textbf{820}, 136557  (2021).}} 






\bibitem{BESIII:2021cvv} M.~Ablikim \textit{et al.} (BESIII Collaboration),
\textcolor{blue}{\href{https://doi.org/10.1103/PhysRevD.105.L011101}{\text{Phys. Rev. D} \textbf{105}, L011101 (2022).}}


\bibitem{BESIII:2023rse}
M.~Ablikim \textit{et al.} (BESIII Collaboration),
\textcolor{blue}{\href{https://doi.org/10.1007/JHEP11(2023)228} {\text{JHEP} \textbf{11}, 228 (2023).}} 





\bibitem{BESIII:2023rwv}
M.~Ablikim \textit{et al.} (BESIII Collaboration),
\textcolor{blue}{\href{https://doi.org/10.1103/PhysRevLett.131.191901}{\text{Phys. Rev. Lett.} \textbf{131}, 191901 (2023).}}

\bibitem{BESIII:2024umc}
M.~Ablikim \textit{et al.} (BESIII Collaboration),
\textcolor{blue}{\href{https://link.springer.com/article/10.1007/JHEP05(2024)022}{\text{JHEP} \textbf{05}, 022 (2024).}}



\bibitem{BESIII:2024ogz}
M.~Ablikim \textit{et al.} (BESIII Collaboration),
\textcolor{blue}{\href{https://link.springer.com/article/10.1007/JHEP07(2024)258}{\text{JHEP} \textbf{07}, 258 (2024).}}

\bibitem{BESIII:2024ues}
M.~Ablikim \textit{et al.} (BESIII Collaboration),,
JHEP \textbf{11}, 062 (2024)
\textcolor{blue}{\href{https://link.springer.com/article/10.1007/JHEP11(2024)062}{\text{JHEP} \textbf{11}, 062 (2024).}}

\bbt{Dobbs:2014ifa-1} S.~Dobbs, K.~K.~Seth, A.~Tomaradze, T.~Xiao and G.~Bonvicini, \href{https://doi.org/10.1103/PhysRevD.96.092004}{Phys.\ Rev.\ D {\bf 96}, 092004 (2017)}.



\bibitem{BCS:+0-}
B. Yan, C. Chen and J.J. Xie
\textcolor{blue}{\href{https://journals.aps.org/prd/abstract/10.1103/PhysRevD.107.076008} {\text{Phys. Rev D} \textbf{107}, 076008 (2023).}} 

\bibitem{Dai:2023vsw}
J.~P.~Dai, X.~Cao and H.~Lenske,
\textcolor{blue}{\href{https://www.sciencedirect.com/science/article/pii/S0370269323005269?via\%3Dihub} {\text{Phys. Lett. B} \textbf{846}, 138192 (2023).}}


\bibitem{vmdmodel}
Y.~Z.~Xu, S.~Y.~Chen, Z.~Q.~Yao \textit{et al.},
\textcolor{blue}{\href{https://doi.org/10.1140/epjc/s10052-021-09673-w} {\text{Eur. Phys. J. C} \textbf{81}, 895 (2021).}} 

\bibitem{Iachello:1972nu}
F.~Iachello, A.~D.~Jackson and A.~Lande,
\textcolor{blue}{\href{https://www.sciencedirect.com/science/article/abs/pii/0370269373902669} {\text{Phys. Lett. B} \textbf{43}, 191--196 (1973).}} 

\bibitem{Iachello:2004aq}
F.~Iachello and Q.~Wan,
\textcolor{blue}{\href{https://journals.aps.org/prc/abstract/10.1103/PhysRevC.69.055204} {\text{Phys. Rev. C} \textbf{69}, L055204 (2004).}} 

\bibitem{Bijker:2004yu}
R.~Bijker and F.~Iachello,
\textcolor{blue}{\href{https://journals.aps.org/prc/abstract/10.1103/PhysRevC.69.068201} {\text{Phys. Rev. C} \textbf{69}, L068201 (2004).}} 

\bibitem{Yang:2019mzq}
Y.~Yang, D.~Y.~Chen and Z.~Lu,
\textcolor{blue}{\href{https://journals.aps.org/prd/abstract/10.1103/PhysRevD.100.073007} {\text{Phys. Rev. D} \textbf{100}, L073007 (2019).}} 

\bibitem{Li:2021lvs}
Z.~Y.~Li, A.~X.~Dai and J.~J.~Xie,
\textcolor{blue}{\href{https://iopscience.iop.org/article/10.1088/0256-307X/39/1/011201} {\text{Chin. Phys. Lett.} \textbf{39}, L011201 (2022).}} 



\bbt{besiii} 
M.~Ablikim \textit{et al.} (BESIII Collaboration),
\textcolor{blue}{\href{https://doi.org/10.1016/j.nima.2009.12.050} {\text{Nucl. Instrum. Meth. A} \textbf{614}, 345-399 (2010).}} 


\bibitem{BEPCII} 
C. Yu \textit{et al.},
\textcolor{blue}{\href{https://doi.org/10.18429/JACoW-IPAC2016-TUYA01} {\text{IPAC 2016} May 8-13 2016.}}


\bibitem{KKMC} 
S. Jadach, B. F. L. Ward and Z. Was,
\textcolor{blue}{ \href{https://doi.org/10.1103/PhysRevD.63.113009} {\text{Phys. Rev. D} \textbf{63}, 113009 (2001).}} 


\bibitem{EVTGEN}
D. J. Lange,
\textcolor{blue}{\href{https://doi.org/10.1016/S0168-9002(01)00089-4} {\text{Nucl. Instrum. Meth. A} \textbf{462}, 152-155 (2001).}}

\bibitem{evtgen2} 
R. G. Ping,
\textcolor{blue}{ \href{https://doi.org/10.1088/1674-1137/32/8/001} {\text{Chin. Phys. C} \textbf{32}, 599 (2008).}}

\bibitem{GEANT4} 
S.~Agostinelli {\it et al.} (GEANT4 Collaboration),
\textcolor{blue}{\href{https://doi.org/10.1016/S0168-9002(03)01368-8} {\text{Nucl. Instrum. Meth. A} \textbf {506}, 250 (2003).}}



\bbt{PDG2020} 
R.~L.~Workman \textit{et al.} (Particle Data Group),
\textcolor{blue}{\href{https://doi.org/10.1093/ptep/ptac097} {\text{PTEP} \textbf{2022}, 083C01 (2022).}}

\bibitem{Lundberg:2009iu}
J.~Lundberg, J.~Conrad, W.~Rolke, and A.~Lopez,
\textcolor{blue}{\href{https://doi.org/10.1016/j.cpc.2009.11.001} {\text{Comput. Phys. Commun.} \textbf{181}, 683-686 (2010).}}
\bbt{SMABCD}
See Supplemental Material at \href{https://journals.aps.org/prd/supplemental/10.1103/PhysRevD.111.L051502/supplement.pdf}{http://link.aps.org/
supplemental/10.1103/PhysRevD.111.L051502} 
for a summary of luminosity,  ISR factor, vacuum polarization factor, detection efficiency, number of signal events,
the Born cross section, the effective form factor, significance at each energy point, which includes Refs.~\cite{BESIII:2015zbz, BESIII:2020eyu}.
\bibitem{BESIII:2015zbz}
M.~Ablikim \textit{et al.} (BESIII Collaboration),
\textcolor{blue}{\href{https://iopscience.iop.org/article/10.1088/1674-1137/40/6/063001} {\text{
Chin. Phys. C} \textbf{40}, 063001 (2016).}}
\bibitem{BESIII:2020eyu}
M.~Ablikim \textit{et al.} (BESIII Collaboration),
\textcolor{blue}{\href{https://iopscience.iop.org/article/10.1088/1674-1137/ac1575} {\text{
Chin. Phys. C} \textbf{45}, 103001 (2021).}}

\bibitem{Kuraev:1985hb}
E.~A.~Kuraev and V.~S.~Fadin,
\textcolor{blue}{\href{https://inspirehep.net/literature/217313}{\text{Sov. J. Nucl. Phys.} \textbf{41}, 466-472 (1985).}}

\bibitem{Jegerlehner:2011ti}
F.~Jegerlehner and R.~Szafron,
\textcolor{blue}{\href{https://doi.org/10.1140/epjc/s10052-011-1632-3} {\text{Eur. Phys. J. C} \textbf{71}, 1632 (2011).}}

\bibitem{Sun:2020ehv}
W.~Sun, T.~Liu, M.~Jing, L.~Wang, B.~Zhong, and W.~Song,
\textcolor{blue}{\href{https://doi.org/10.1007/s11467-021-1085-6} {\text{Front. Phys. (Beijing)} \textbf{16}, 64501 (2021).}} 


\bibitem{symm:wang} 
X.~F.~Wang and G.~S.~Huang,
\textcolor{blue}{\href{https://doi.org/10.3390/sym14010065} {\text{Symmetry} \textbf{2022}, 14, 65.}} 

%\bibitem{symm:xia} L.~Xia, C.~Rosner, Y.~D.~Wang, X.~R.~Zhou, F.~E.~Maas, R.~B.~Ferroli, H.~M.~Hu and G.~S.~Huang, 
%\textcolor{blue}{\href{https://doi.org/10.3390/sym14020231} {\text{Symmetry} \textbf{2022}, 14, 231.}} 

\bbt{Baldini}
R.~Baldini, S.~Pacetti, A.~Zallo and A.~Zichichi,
\textcolor{blue}{\href{https://doi.org/10.1140/epja/i2008-10716-1} {\text{Eur. Phys. J. A} \textbf{39}, 315-321 (2009).}} 

\bbt{Arbuzov} 
A.~B.~Arbuzov and T.~V.~Kopylova,
\textcolor{blue}{\href{https://doi.org/10.1007/JHEP04(2012)009} {\text{JHEP} \textbf{04}, 009 (2012).}} 


\bibitem{BESIII:2015qfd}
M.~Ablikim \textit{et al.} (BESIII Collaboration),
\textcolor{blue}{\href{https://doi.org/10.1088/1674-1137/39/9/093001} {\text{Chin. Phys. C} \textbf{39}, 093001 (2015).}} 

\bibitem{BESIII:2022dxl}
M.~Ablikim \textit{et al.} (BESIII Collaboration),
\textcolor{blue}{\href{https://doi.org/10.1088/1674-1137/ac80b4} {\text{Chin. Phys. C} \textbf{46}, 113002 (2022).}} 
\bibitem{BESIII:2022ulv}
M.~Ablikim \textit{et al.} (BESIII Collaboration),
\textcolor{blue}{\href{https://doi.org/10.1088/1674-1137/ac84cc} {\text{Chin. Phys. C} \textbf{46}, 113003 (2022).}} 







\bibitem{BESIII:2016ssr}
M.~Ablikim \textit{et al.} (BESIII Collaboration),
\textcolor{blue}{\href{https://journals.aps.org/prd/abstract/10.1103/PhysRevD.93.072003} {\text{Phys. Rev. D} \textbf{93}, 072003 (2016).}} 

\bibitem{BESIII:2016nix}
M.~Ablikim \textit{et al.} (BESIII Collaboration),
\textcolor{blue}{\href{https://www.sciencedirect.com/science/article/pii/S0370269317303222?via\%3Dihub} {\text{Phys. Lett. B} \textbf{770}, 217-225 (2017).}} 


\bibitem{BESIII:2019dve}
M.~Ablikim \textit{et al.} (BESIII Collaboration),
\textcolor{blue}{\href{https://journals.aps.org/prd/abstract/10.1103/PhysRevD.100.051101} {\text{Phys. Rev. D} \textbf{100}, 051101 (2019).}} 





\bibitem{BESIII:2022mfx}
M.~Ablikim \textit{et al.} (BESIII Collaboration),
\textcolor{blue}{\href{https://link.springer.com/article/10.1007/JHEP06(2022)074} {\text{JHEP} \textbf{06}, 74 (2022).}} 

\bibitem{BESIII:2022lsz}
M.~Ablikim \textit{et al.} (BESIII Collaboration),
\textcolor{blue}{\href{https://journals.aps.org/prd/abstract/10.1103/PhysRevD.106.L091101} {\text{Phys. Rev. D} \textbf{106}, L091101 (2022).}} 

\bibitem{BESIII:2023lkg}
M.~Ablikim \textit{et al.} (BESIII Collaboration),
\textcolor{blue}{\href{https://journals.aps.org/prd/abstract/10.1103/PhysRevD.108.L011101} {\text{Phys. Rev. D} \textbf{108}, L011101 (2023).}} 


\bibitem{BESIII:2023euh}
M.~Ablikim \textit{et al.} (BESIII Collaboration),
\textcolor{blue}{\href{https://link.springer.com/article/10.1007/JHEP10(2023)081} {\text{JHEP} \textbf{10}, 081 (2023)}} 
\textcolor{blue}{\href{https://link.springer.com/article/10.1007/JHEP12(2023)080} {\text{[erratum: JHEP} \textbf{12}, 080 (2023)].}} 





\bibitem{BESIII:lambda_rec_eff}
M.~Ablikim \textit{et al.} (BESIII Collaboration),
\textcolor{blue}{\href{https://journals.aps.org/prd/abstract/10.1103/PhysRevD.108.L031106} {\text{Phys. Rev. D} \textbf{108}, L031106 (2023).}} 





\bibitem{Zhu:2008ca}
Y.~S.~Zhu,
\textcolor{blue}{\href{https://doi.org/10.1088/1674-1137/32/5/007} {\text{Chin. Phys. C} \textbf{32}, 363 (2008).}} 


\end{thebibliography}
\end{document}